\documentclass[runningheads]{llncs}

\RequirePackage{fix-cm}

\usepackage[noadjust]{cite}

\usepackage{graphicx}
\usepackage{epsfig}
\usepackage{dsfont}
\usepackage{mathtools}
\usepackage{amssymb}
\usepackage{boldline,multirow}
\usepackage{diagbox}
\usepackage{ulem}
\usepackage{color}
\usepackage{graphicx}
\usepackage{subfigure}
\usepackage{bbding}
\usepackage[colorlinks,
linkcolor=red,
anchorcolor=blue,
citecolor=blue
]{hyperref} 
\begin{document}
	\title{Learning Incrementally to Segment Multiple Organs in a CT Image} 
	%
	%
	%
	\author{Pengbo Liu\inst{2} \and
		Xia Wang\inst{3}\and
		Mengsi Fan\inst{3}\and
		Hongli Pan\inst{3}\and
		Minmin Yin\inst{3}\and
		Xiaohong Zhu\inst{3}\and
		Dandan Du\inst{3}\and
		Xiaoying Zhao\inst{3}\and
		Li Xiao\inst{2} \and
		Lian Ding\inst{4}\and
		Xingwang Wu\inst{3}\and
		S. Kevin Zhou\inst{1,2}}
	
	\authorrunning{Pengbo Liu et al.}
	%
	\institute{School of Biomedical Engineering \& Suzhou Institute for Advanced Research Center for Medical Imaging, Robotics, and Analytic Computing \& Learning (MIRACLE) University of Science and Technology of China, Suzhou 215123, China \and
			Key Lab of Intelligent Information Processing of Chinese Academy of Sciences (CAS),
			Institute of Computing Technology, CAS, Beijing, 100190, China	\email{liupengbo2019@ict.ac.cn} \and
			The First Affiliated Hospital of Anhui Medical University, Anhui, China	 \and
			Huawei Cloud Computing Technology Co. Ltd, China
		}
	
	
	\maketitle              
	\begin{abstract}
		
		There exists a large number of datasets for organ segmentation, which are partially annotated and sequentially constructed. A typical dataset is constructed at a certain time by curating medical images and annotating the organs of interest. In other words, new datasets with annotations of new organ categories are built over time. 
		To unleash the potential behind these partially labeled, sequentially-constructed datasets, we propose to incrementally learn a multi-organ segmentation model. In each incremental learning (IL) stage, we lose the access to previous data and annotations, whose knowledge is assumingly captured by the current model, and gain the access to a new dataset with annotations of new organ categories, from which we learn to update the organ segmentation model to include the new organs.
		While IL is notorious for its `catastrophic forgetting' weakness in the context of natural image analysis, we experimentally discover that such a weakness mostly disappears for CT multi-organ segmentation. 
		To further stabilize the model performance across the IL stages,
		we introduce a \textit{light memory module} and some loss functions to restrain the representation of different categories in feature space, aggregating feature representation of the same class and separating feature representation of different classes.
		Extensive experiments on five open-sourced datasets are conducted to illustrate the effectiveness of our method. 
		
	\end{abstract}
	
	\section{Introduction}
	\label{intro}
	
	While most natural image datasets~\cite{imagenet,coco} are completely labeled for common categories, 
	fully annotated medical image datasets are scarce, especially for a multi-organ segmentation (MOS) task \cite{zhou2019handbook} that requires pixel-wise annotations, as constructing such a dataset requires professional knowledge of different anatomical structures~\cite{zhou2019handbook, zhou2021review}.
	Fortunately, there exist many partially labeled datasets \cite{MSD, matlas, kits19_url3} for organ segmentation. 
    Another dimension associated with these datasets is that they are constructed sequentially at different sites.
	Our goal is to train \textbf{a single multi-organ segmentation model from partially labelled, sequentially constructed datasets}.
	
\begin{figure}[t]
		\centering
		\includegraphics[width=0.7\textwidth]{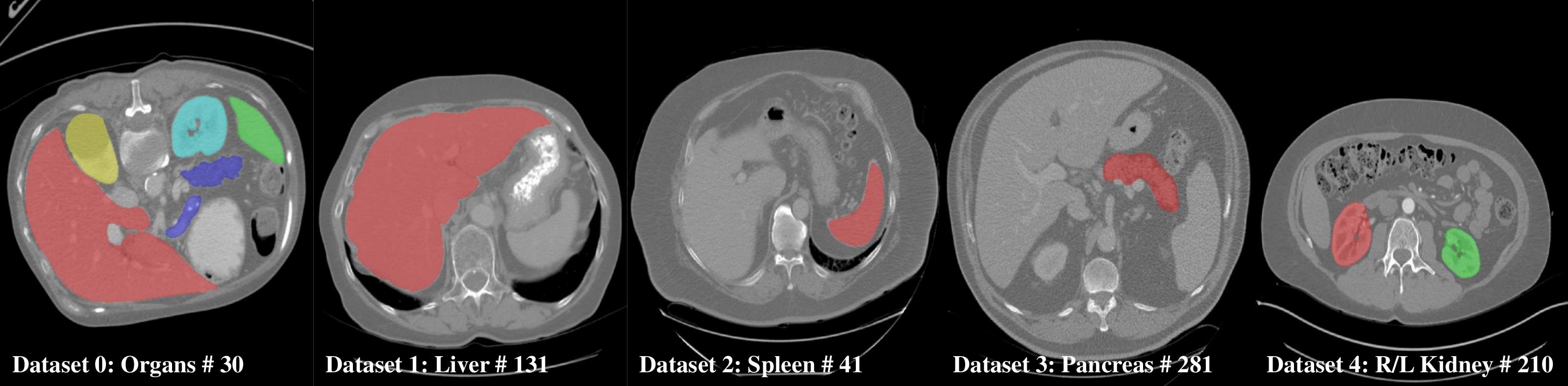}
		\caption{Number of cases in different partially labeled datasets for different tasks.
		} \label{FigOrgans}
	\end{figure}

	To achieve such a goal, we have to address two issues. (i) The first issue arising from \textit{partial labeling} is  \underline{knowledge conflict}, that is, labels in different datasets have conflicts, \textit{e.g.}, the liver is marked as foreground in Dataset 1 but as background in Datasets 2-4, as shown in Fig.~\ref{FigOrgans}. (ii) The second issue arising from \textit{sequential construction} is \underline{data availability}, that is, the datasets are not simultaneously available for learning. What could be even worse is that, due to \underline{security concern}, these datasets are not allowed to be transferred across the border of the curating institutes; only the model parameters are sharable.
	
	
	There has been some emerging research~\cite{cvpr2021dodnet,PIPO,PaNN,ShiMargExc} that successfully handles knowledge conflict and trains a single model from pooled datasets for improved performance in multi-organ segmentation, proving that the unlabeled data in partially labeled datasets is also helpful for learning. 
	However, these approaches conduct model learning in a batch model based and hence unable to be applied to deal with sequential construction. To deal with both issues, we hereby propose a novel multi-organ segmentation approach based on \textit{the principle of incremental learning (IL)}, which is a staged learning method that has an access to the data available at current learning stage, while losing the access to the data available in previous stages.

	Our main contributions are summarized as below:
	\begin{itemize}
		\item We make the first attempt in the literature to merge partially labeled datasets in medical image scenario using IL method, addressing the issues of knowledge conflict and data availability, and possibly security concern. 
		\item To combat the `catastrophic forgetting' problem that commonly plagues IL, we introduce a light memory module to store the prototypical representation of different organ categories and corresponding loss functions to make different organs more distinguishable in feature space.
		\item Our extensive experiments on five open-source organ datasets achieve comparable performance to state-of-the-art (SOTA) batch methods which can access all datasets in training phase, unleashing the great potential of IL in multiple organ segmentation.
	\end{itemize}
	
\section{Related work}
	
\noindent\textbf{MOS with Partially Labelled Datasets.} 
Zhou \textit{et al.}~\cite{PaNN} learn a segmentation model in the case of partial labeling by adding a prior-aware loss in the learning objective to match the distribution between the unlabeled and labeled datasets.
In \cite{PIPO}, first multi-scale features at various depths are hierarchically incorporated for image segmentation and then a unified segmentation strategy is developed to train three separate datasets together, and finally multi-organ segmentation is achieved by learning from the union of partially labeled and fully labeled datasets. 
Zhang \textit{et al.} \cite{cvpr2021dodnet} propose a dynamic on-demand network (DoDNet) that learns to segment multiple organs and tumors on partially labeled datasets, which embedded dynamically generated filter by a task encoding module into an encoder-decoder architecture. Shi \textit{et al.}~\cite{ShiMargExc} encode knowledge from different organs into a single multi-class segmentation model by introducing two simple but effective loss functions, \textit{Marginal} loss and \textit{Exclusion} loss. 
	
		

\noindent\textbf{Incremental Learning.} IL has been studied for object recognition~\cite{lwf,kirkpatrick2017overcoming,icarl2017,LUCIR2019,meta2020} and detection~\cite{9035099,8288446,shmelkov2017incremental}, also segmentation~\cite{sensing2019,ILT,MiB,ozdemir2019extending}. The main challenge in IL is the so-called {`catastrophic forgetting'}~\cite{mccloskey1989catastrophic}: how to keep the performance on old classes while learning new ones?
Methods based on parameter isolation~\cite{rusu2016progressive, xu2018reinforced} and data replay~\cite{icarl2017, lopez2017gradient} are all with limited scalability or {privacy issues}. 
Regularization based method is the most ideal direction in IL community.
In natural image segmentation, Cermelli et al.~\cite{MiB} solved knowledge conflicts existing in other IL methods~\cite{lwf,ILT} by remodeling old and new categories into background in loss functions,
achieving a performance improvement.
In 2D medical image segmentation, Ozdemir and Goksel~\cite{ozdemir2019extending} made some attempts using the IL methods used in natural images directly, with only two categories, and it mainly focuses on verifying the possibility of transferring the knowledge learned in the first category with more images to a second category with less images. In this paper, we apply IL to multiple organ segmentation for the first time.

	\section{Method}

	\subsection{IL for MOS}
\begin{figure}[t]
	\centering
	\includegraphics[width=0.7\textwidth]{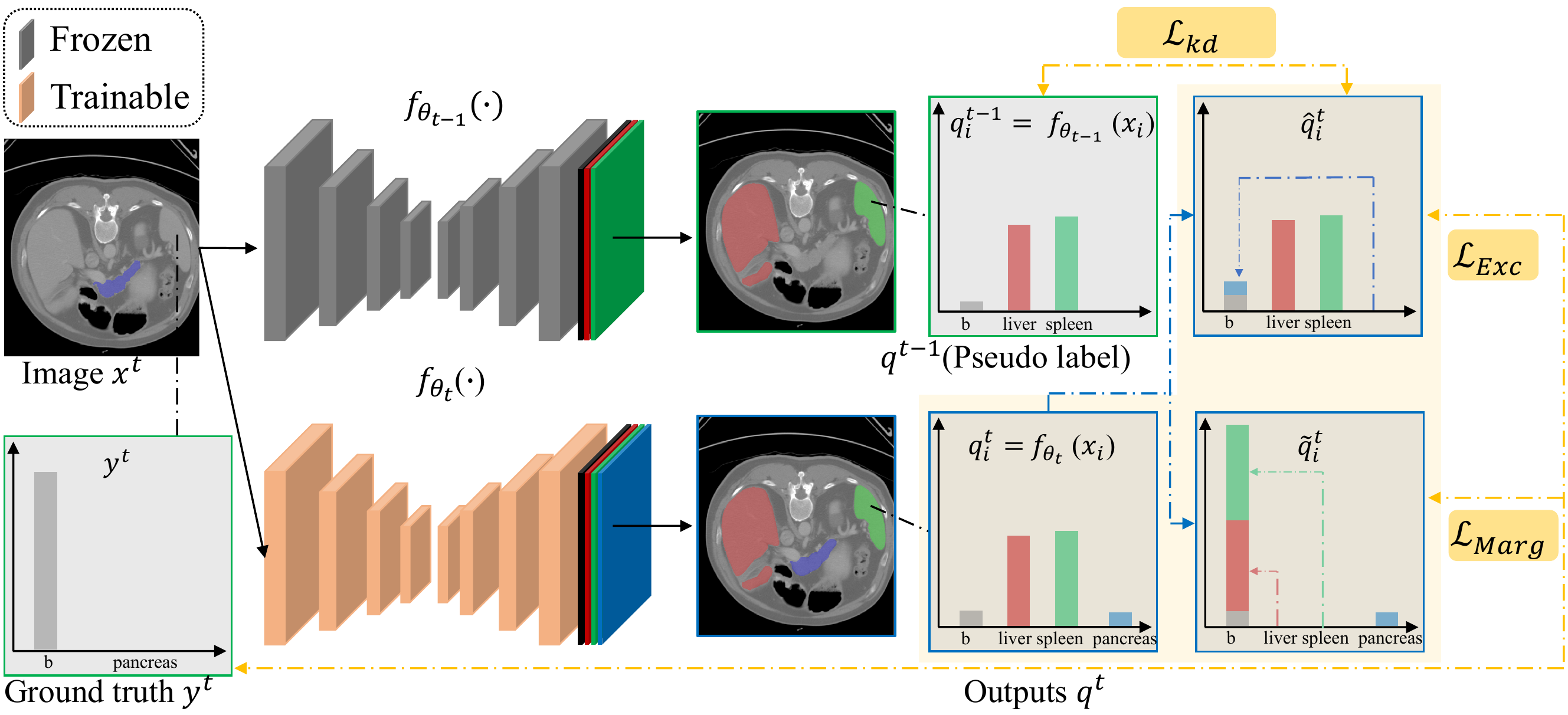}
	\caption{Overview of the $t^{th}$ stage of IL in multi-organ segmentation.} \label{Overview}
	\end{figure}

	\textbf{Framework of IL. }The overview of the $t^{th}$ stage of IL in our method is shown in Fig.~\ref{Overview}. Given a pair of 3D input image and ground truth, $\{x^t, y^t\}\in \{\mathcal{X}^t, \mathcal{C}^t\}$, 
	we firstly process $x^t$ by the model in current stage, $f_{\theta_t}(\cdot)$ with trainable parameters $\theta_t$, getting the output $q^t=f_{\theta_t}(x^t)$. And we assume that each image $x^t$ is composed by a set of voxels $x^t_i$ with constant cardinality $|\mathcal{I} | = N$. 
	The whole label space $\mathcal{Y}^t$ cross all $t$ stages is expanded from $\mathcal{Y}^{t-1}$ with new classes added in current stage ($\mathcal{C}^t$), $\mathcal{Y}^t=\mathcal{Y}^{t-1} \cup \mathcal{C}^t=\mathcal{C}^1\cup ...\cup \mathcal{C}^t$. 
	Note that the annotations of the old categories $\mathcal{Y}^{t-1}$ will be inaccessible in the new stage under ideal IL settings. For preserving the knowledge of old categories in regularization based method, we process $x^t$ by the saved old model $f_{\theta_{t-1}}(\cdot)$ with frozen parameters $\theta_{t-1}$ and get $q^{t-1}=f_{\theta_{t-1}}(x^t)$ as the pseudo label. 
	Knowledge distillation loss, $\mathcal{L}_{kd}$, is introduced in IL setting to keep old knowledge learned from previous stages. Trainable $\theta_t$ in the $t^{th}$ stage is expanded from $\theta_{t-1}$ with $\Theta_{t}$ to segment new categories, $\theta_t=\theta_{t-1}\cup\Theta_{t}$. 
	
	\noindent\textbf{Avoiding Knowledge Conflict in IL.} The structures of old classes in $\mathcal{X}^t$,  are marked as background in $\mathcal{C}^t$. And the new structures also do not exist in $\mathcal{Y}^{t-1}$, that is new structures are marked as background in pseudo label. If we directly use $q^t$ to compute segmentation loss for new classes, and knowledge distillation loss for old classes, these conflicts between prediction and ground truth break the whole training process. So referring to marginal loss in MargExc~\cite{ShiMargExc}, we modify the prediction $q^t$ to $\hat{q}^t$ and $\tilde{q}^t$, as shown in Fig.~\ref{Overview} and Eqs.~(\ref{margin}) and (\ref{margin_2}). 
	
	\begin{align}\label{margin}
	\hat{q}^t_{i,j}&=\left\{
				\begin{array}{lp{8mm}<{\centering}l}
				\text{exp}(q^t_{i,b}+\sum_{c\in \mathcal{C}^{t}}q^t_{i,c})/\sum_{c\in \mathcal{Y}^{t}\cup b}\text{exp}(q^t_{i,c}) & &\textit{if } j = b\\
			\text{exp}(q^t_{i,j})/\sum_{c\in \mathcal{Y}^t\cup b}\text{exp}(q^t_{i,c})&& \textit{if } j \in \mathcal{Y}^{t-1} 
				\end{array} \right. \\
	\tilde{q}^t_{i,j}&=\left\{
					\begin{array}{lp{8mm}<{\centering}l} 
				\text{exp}(q^t_{i,b}+\sum_{c\in \mathcal{Y}^{t-1}}q^t_{i,c})/\sum_{c\in \mathcal{Y}^{t}\cup b}\text{exp}(q^t_{i,c}) & &\textit{if } j = b\\
			0 && \textit{if } j \in \mathcal{Y}^{t-1} \\
			\text{exp}(q^t_{i,j})/\sum_{c\in \mathcal{Y}^{t}\cup b}\text{exp}(q^t_{i,c}) & & \textit{if }  j \in \mathcal{C}^t
			\end{array} \right.\label{margin_2}
	\end{align}
	Then the probability of classes not marked in ground truth or pseudo label will not be broken during training.
	

	\subsection{Memory Module}
\begin{figure}[t]
		\centering
		\includegraphics[width=0.9\textwidth]{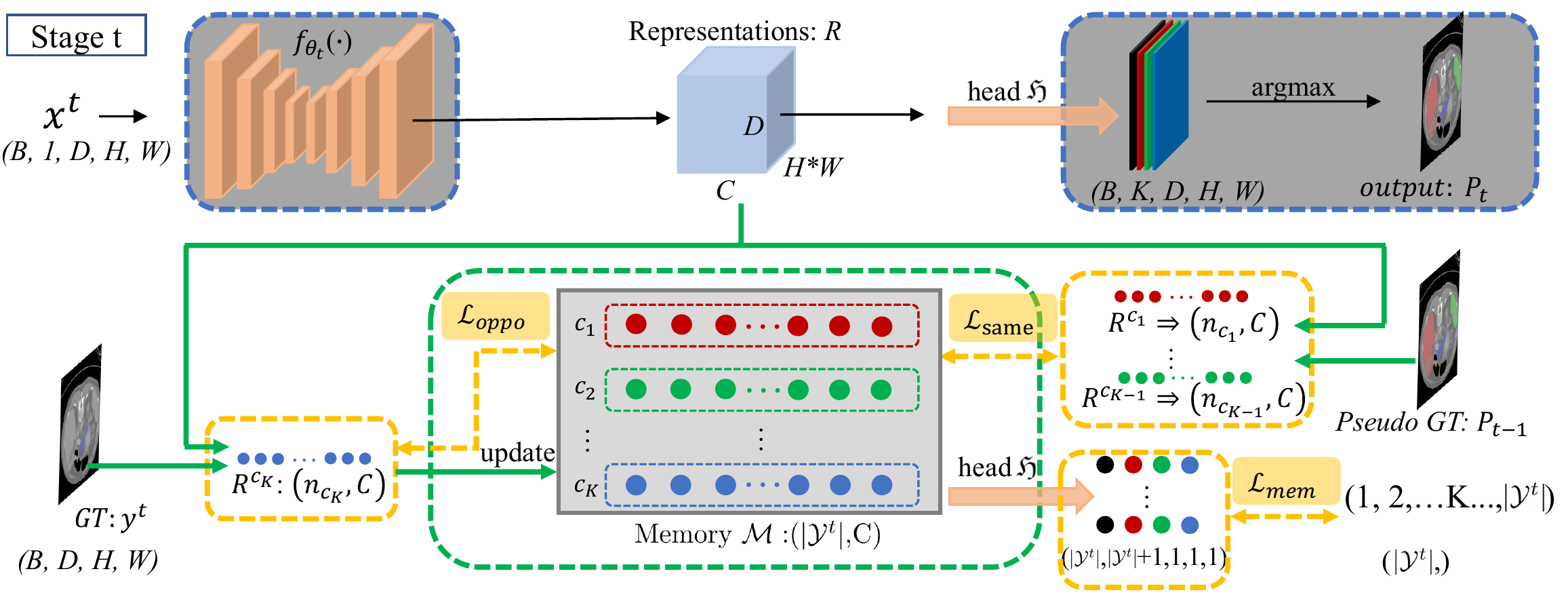}
		\caption{Diagram of the memory module $\mathcal{M}$ in feature space between decoder and segmentation head. Based on label, we can take $n_{c_K}$ voxels' representation from $\mathcal{R}$, $\mathcal{R}^{c_K}$:($n_{c_K}$, C), to update $\mathcal{M}$ or calculate loss function.}\label{memorymodule}
	\end{figure}

	As shown in Fig.~\ref{memorymodule}, representation $\mathcal{R}$ is feature maps out of decoder, with shape of C$\times$D$\times$H$\times$W, where C means the number of channels in $\mathcal{R}$. 
	To further mitigate `knowledge forgetting' in IL setting, we introduce a light memory module $\mathcal{M}$ of size $\left|\mathcal{Y}^t\right|\times$C in feature space between decoder and segmentation head, $\mathfrak{H}$, to remember the representation of each class. The size of $\mathcal{M}$ is updated by more $\left|\mathcal{C}^t\right|\times$C  on $\left|\mathcal{Y}^{t-1}\right|\times$C after the $t^{th}$ stage. Then based on $\mathcal{M}$ we can add some constraints in feature space to improve the IL learning progress. 
	
	During training of each stage, with the position supplied by ground truth, we can acquire the voxel representation of corresponding new organs in feature map $\mathcal{R}$. Then new class $c$ in $\mathcal{M}$ can be updated via moving average after each iteration:
	\begin{equation}\label{M}
	\begin{aligned}
	\mathcal{M}^c_{k} = (1-m_{k})\cdot\mathcal{M}^c_{k-1} + m_{k}\cdot\mathcal{R}^c_{k},~~
	m_k = \frac{9m_0}{10}\cdot(1-\frac{k}{K})^p+\frac{m_0}{10},
	\end{aligned}
	\end{equation}
	where $m$ is the momentum, $k$ denotes the current number of iterations, and $K$ is the total number of iterations of training. $p$ and $m_0$ are set as 0.9 empirically. After each stage of training ends, the mean representation of new organ of category $c$ in that stage is saved into the memory $\mathcal{M}$ as $\mathcal{M}^c$.
	
	When we have $\mathcal{M}$ to save the mean representation of each class, we can introduce more regularization to constrain the learning of feature space. In this paper, we introduce $l_{mem}$, $l_{same}$ and $l_{oppo}$:
	
	
	
	

	    \begin{align}
	    l_{mem}=&\mathcal{L}_{ce}(\mathfrak{H}(reshape(\mathcal{M})), range(1,\left|\mathcal{Y}^{t}\right|+1))\label{lmem}\\
	    l_{same}=&\sum_{c_0\in {\mathcal{Y}^{t-1}}} \mathcal{L}_{cos}(\mathcal{M}^{c_o}, \mathcal{R}^{c_o},1)\label{lsame}\\
	    l_{oppo}=&\sum_{c_n\in \mathcal{C}^t}(\mathcal{L}_{cos}(\mathcal{R}^{b}, \mathcal{R}^{c_n},-1)+\sum_{c_0\in {\mathcal{Y}^{t-1}}} \mathcal{L}_{cos}(\mathcal{M}^{c_o}, \mathcal{R}^{c_n},-1))\label{loppo}
	    \end{align}

	In Eq.~(\ref{lmem}), $reshape$ is used to change $\mathcal{M}$ to the size of $\left|\mathcal{Y}^t\right| \times C \times 1 \times 1\times 1$, which can be regarded as $\left|\mathcal{Y}^t\right|$ voxels belong to  $\left|\mathcal{Y}^t\right|$ classes. $range(1,\left|\mathcal{Y}^{t}\right|+1)$ can be seen as corresponding ground truth. Through the shared segmentation head $\mathfrak{H}$, features of classes in current stage are going to center around the mean representation in $\mathcal{M}$. Through $l_{mem}$, we constrain the learned feature of different classes in different stages more stable. The mean representation of old classes are treated as a kind of replay without privacy concerns.
	In Eqs.~(\ref{lsame}) and (\ref{loppo}), $c_o$ and $c_n$ refer to old and new classes, respectively, and $b$ means background. Using Cosine Embedding Loss, $\mathcal{L}_{cos}$, we can explicitly restrain the feature of old class close to $\mathcal{M}^{c_o}$, and the feature of new class away from all $\mathcal{M}^{c_o}$.

	\section{Experiments}
	
	\subsection{Setup}

	\textbf{Datasets and preprocessing.} 
	To compare with our base method, MargExc~\cite{ShiMargExc}, we choose the same five organs and datasets in our experiments, including liver, spleen, pancreas, right kidney and left kidney. In addition, we find three more independent datasets for testing to give a comprehensive evaluation. The details of these datasets are shown in Table~\ref{summaryofdatasets}.
	
	We preprocess all datasets to a unified spacing (2.41, 1.63, 1.63) and normalize them with mean and std of 90.9 and 65.5 respectively. We respectively split five training datasets into 5 folds and randomly select one fold as validation set. 
	For our main IL setting, five organs are learned in four stages: liver (F+$P_1$)$\rightarrow$ spleen (F+$P_2$)$\rightarrow$ pancreas (F+$P_3$)$\rightarrow$ R/L kidney (F+$P_4$). The annotations of different organs in dataset F are used separately in our IL setting.

	\noindent\textbf{Implementation details.} We implement our experiments based on 3D lowres version of nnU-Net\footnote{\url{github.com/mic-dkfz/nnunet}}~\cite{FabianNNUnet_nm} and also refer to MONAI\footnote{\url{https://monai.io/}}
	during our algorithm development. 
	The patch-size and batch-size are set as (80, 160, 128) and 2, respectively, in our experiments. We train the network with the same optimizer and learning rate policy as nnU-Net for 350 epochs. The initial learning rate of the first stage and followed stages are set to 3e-4 and 15e-5.
	
	\begin{table}[t]
		\centering
		\caption{A summary of five benchmark datasets used in our experiments. [T] means there are tumor labels in original dataset and we merge them into corresponding organs.}\label{summaryofdatasets}
		\newsavebox{\tablebox}
		\begin{lrbox}{\tablebox}
			\begin{tabular}{c|l|c|c|c|c|c}
				\hline
				Phase&Datasets &  Modality &  \# of labeled volumes & Annotated organs & Mean spacing (z, y, x) & Source\\
				\hline
				\multirow{6}{1.7cm}{\centering $Training$\\$\& Val$}&Dataset0 (F) 	 & CT & 30 	& Five organs 	& (3.0, 0.76, 0.76) &Abdomen in \cite{matlas}\\
				&Dataset1 ($P_1$) & CT & 131 & Liver [T]		& (1.0, 0.77, 0.77) 	&Task03 in ~\cite{MSD}\\
				&Dataset2 ($P_2$) & CT & 41 	& Spleen 			& (1.6, 0.79, 0.79) &Task09 in ~\cite{MSD}\\
				&Dataset3 ($P_3$) & CT & 281 & Pancreas [T] 	& (2.5, 0.80, 0.80) 		&Task07 in ~\cite{MSD}\\
				&Dataset4 ($P_4$) & CT & 210 & L\&R Kidneys [T] & (0.8, 0.78, 0.78) 		&KiTS~\cite{kits19_url3}\\
				\cline{2-7}	
				&All				 & CT & 693 & Five organs & (1.7, 0.79, 0.79) & -\\
				\hline\hline
				\multirow{3}{1.7cm}{\centering $Testing$}&CLINIC & CT & 107 & Five organs & (1.2, 0.74, 0.74) 		& Private\\
				&Amos & CT & 200 & Five organs & (5.0, 0.74, 0.74)	&Temporarily private\\
				&Pan  & CT & 56 &Five organs  &  (2.6, 0.82, 0.82)		&FLARE 21~\cite{flare21} \\
				\hline
			\end{tabular}
		\end{lrbox}
		\scalebox{0.65}[0.65]{\usebox{\tablebox}}
	\end{table}

	\noindent\textbf{Baseline methods.}
	Intuitively, we train a 5-class segmentation model $\phi_F$ on dataset F directly. And to use more partially labeled datasets, we train 4 models separately for different organs, too, i.e., $\phi_{F+P_*}$. To simulate different organs are collected sequentially, simple fine-tuning (FT) and some SOTA IL methods (LwF~\cite{lwf}, ILT~\cite{ILT} and MiB~\cite{MiB}) are also implemented. In the end, to evaluate our performance in actual usage scenarios, we compare our method to the upper bound results from MargExc~\cite{ShiMargExc}. Since we also use the marginal loss in IL, we call our method MargExcIL.
	
	
	\noindent\textbf{Performance metrics.} We use Dice coefficient (DC) and 95$^{th}$ percentile Hausdorff distance (HD95) to evaluate results.

	\subsection{Results and Discussions}
	\label{resultandDis}
	
	\begin{table}[t]
		\centering
		\caption{In the last stage($4^{th}$), the DC and HD95 of the segmentation results of different methods. MargExc~\cite{ShiMargExc} is the upper bound method training all datasets in the meantime. `-' means no result.}\label{wholeresult}
		\begin{lrbox}{\tablebox}
			\begin{tabular}{c|l|rr|rr|rr|rr|rr|rr}
				\hline
				\multirow{3}{3cm}{\centering Training form}&\multirow{3}*{\centering \diagbox{Methods}{Organs}} & \multicolumn{12}{c}{DC/HD95 (${F+P_i}$)}\\
				\cline{3-14}
				& &  \multicolumn{2}{c}{Liver} & \multicolumn{2}{c}{Spleen} & \multicolumn{2}{c}{Pancreas} & \multicolumn{2}{c}{R Kidney} &\multicolumn{2}{c}{L Kidney} & \multicolumn{2}{c}{Mean}\\
				\cline{3-14}
				& & DC & HD & DC & HD & DC & HD & DC & HD & DC & HD & DC & HD\\
				\hline\hline
				\multirow{5}{1.7cm}{\centering $Trained$\\$separated$}&$\phi_{F}$ (Five organs) 		& .953&10.28  & .953&1.93 & .721&8.25 & .895&5.82 & .839&13.41 & .872&7.94 \\
				\cline{2-14}
				&$\phi_{F+P_1}$ (Liver) 		& .967&5.89  & -&- & -&- & -&- & -&- & \multirow{4}{*}{.936} & \multirow{4}{*}{8.25} \\
				&$\phi_{F+P_2}$ (Spleen)  		& -&-  & .954&20.20 & -&- & -&- & -&- & & \\
				&$\phi_{F+P_3}$ (Pancreas)  	& -&-  & -&- & .842&5.13 & -&- & -&- &  &\\
				&$\phi_{F+P_4}$ (Kidneys)  	& -&-  & -&- & -&- & .968&5.18 & .950&4.86 & & \\
				\hline\hline
				
				\multirow{4}{1.7cm}{\centering $One$\\$model$}&FT				& .000&-  & .000&- & .000&- & .970&6.502 & .963&2.018 & .387&- \\
				&LwF~\cite{lwf} 		& .001&190.33  & .906&2.22 & .792&5.91 & .966&7.70 & .948&7.22& .723&42.68 \\
				&ILT~\cite{ILT} 	& .000&170.77  & .914&2.05 & .772&8.40 & .969&1.41 & .948&4.06 & .721&37.34 \\

				&MiB~\cite{MiB} 		& .966&6.76  & .961&1.26 & .817&6.56 & .966&3.77 & .946&7.22 & .931&5.11 \\ 
				\hline
				$Ours$&MargExcIL 		& .965&7.98  & .962&1.30 & .835&5.51 & .968&1.40 & .959&2.37 & \textcolor{red}{.938}&\textcolor{red}{3.71} \\
				
				\hline\hline
				Upper bound&MargExc~\cite{ShiMargExc} 	& .962&7.01  & .965&1.15 & .848&4.83 & .969&1.39 & .965&3.96& \textbf{.942}&\textbf{3.67} \\	
				\hline
			\end{tabular}
		\end{lrbox}
		\scalebox{0.62}[0.62]{\usebox{\tablebox}}
	\end{table}
	

	\subsubsection{Comparison with baseline methods} 
	In IL setting, performance of batch learning of all categories is seen as the upper bound for comparison. Because joint learning can access all knowledge at the meantime, it is possible to fit the distribution of the whole dataset. We regard MargExc~\cite{ShiMargExc} as the counterpart batch method in MOS, which obtains the DC of 0.942 and HD95 of 3.67 when training all five training datasets together, as in Table~\ref{wholeresult}.
	
	When we do not aggregate these partially labeled data together, there are some limitations in performance. 
	The 5-class segmentation model $\phi_F$ only trained on small scale `fully' annotated dataset F, can not generalize well to all validation datasets due to the scale of the dataset F. The metrics of DC and HD95 are all much worse than upper bound. 
	When we train four models, $\phi_{F+P_*}$, one model per organ segmentation task trained on corresponding datasets (F+$P_*$), then all datasets can be used. We can get much better performance than $\phi_F$ on DC metric, but also bad HD95 metric. Higher HD95 means more false positive predictions out of our trained models. Furthermore, training separately is also poor in scalability and efficiency when the categories grow in the future. 
	
	When we aggregate these partially labeled datasets together sequentially, the most intuitive method FT is the worst. It has no preservation of the old knowledge because there is no restraint for it. LwF~\cite{lwf} and ILT~\cite{ILT} perform better than FT, but `knowledge conflict' limits the performance of LwF and ILT when the stage of IL is more than 3, \textit{i.e.}, the liver knowledge in the 1$^{st}$ stage can not be kept in the 4$^{th}$ stage. We check the output of the models trained via LwF and ILT, finding that old organs' logit is overwhelmed by the logit of background as the training stages progress. MiB~\cite{MiB} can get a good result compared with LwF and ILT because of remodeling background and foreground in training phase, thus avoiding the `knowledge conflict' problem. 
	
	MargExc~\cite{ShiMargExc} also solves the `knowledge conflict' problem, which is the most harmful factor in aggregating partially labeled dataset. Based on MargExc, our MargExcIL also performs well on DC and HD95, better than all other methods, \textit{e.g.} MiB or the models trained separated for all organs, approaching upper bound result (DC: 0.938 vs 0.942 \& HD95: 3.71 vs 3.67). In `\textit{Model$_{S3}$ on Testing Datasets}' part in Table~\ref{memoryresult}, MargExcIL even performs better than MargExc~\cite{ShiMargExc}. These results prove IL might have a practical potential in clinical scenario.

	\noindent\textbf{Effectiveness of memory module} 
	In Table~\ref{memoryresult}, 
	we also show the results of the 4 intermediate stages of our method, in `\textit{Model$_{S*}$ on Validation Sets}' part. `$(woMem)$' means our method without memory module and corresponding loss functions. `$_{swin}$' means that we modify encoder of our network designed by nnUNet to Swin Transformer~\cite{liu2021swin}, which can also assist in proving the effectiveness of our memory module.
	Without memory module, we can also obtain the same level performance in last stage, but it's \textit{not stable in the middle stages}, e.g., liver's HD95 get worse dramatically in stage 2 and stage 3. This uncertainty factor in our IL system is not acceptable. We believe that this phenomenon is caused by the variation in the image distribution or field-of-view (FOV) in different datasets. Our memory module stores a prior knowledge of old class to stabilize the whole IL system. Compared with MargExc~\cite{ShiMargExc}, we also achieve a comparable performance. 

	\begin{table}[t]
		\centering
		\caption{The DC and HD95 of the segmentation results. The best and second result is shown in \textbf{bold} and \textcolor{red}{red}. S$*$ means $*^{th}$ stage in IL setting. `-' means \textbf{No Access} to the classes in that stage. }\label{memoryresult}
		\begin{lrbox}{\tablebox}
			\begin{tabular}{l|l|rr|rr|rr|rr||rr|rr|rr}
				\hline
				\multirow{3}{2cm}{\centering Setting}&\multirow{3}*{\centering \diagbox{Organs}{DC/HD95}}&\multicolumn{8}{c||}{Model$_{S*}$ on Validation Sets}&\multicolumn{6}{c}{Model$_{S3}$ on Testing Datasets}\\
				\cline{3-16}
				&  &  \multicolumn{2}{c}{S0} & \multicolumn{2}{c}{S1} & \multicolumn{2}{c}{S2} & \multicolumn{2}{c}{S3} & \multicolumn{2}{c}{CLINIC} & \multicolumn{2}{c}{AMOS} & \multicolumn{2}{c}{Pan}   \\
				\cline{3-16}
				& & DC & HD & DC & HD & DC & HD & DC & HD & DC & HD & DC & HD & DC &  HD\\
				\hline\hline
				\multirow{6}{2cm}{\centering $MargExc-$\\$IL_{swin}$\\$(woMem)$}	    &Liver 			& .965&5.51 & .959&20.10  &.958&20.86  &.957&7.24 &.971&3.33 &.937&10.68 & .977&2.00  \\
				&Spleen 		& -&- &.958&2.11   &.960&2.12  &.962&1.19  &.953&5.88 &.865&6.93 &.965&1.66  \\
				&Pancreas 		&-&-  & -&-  & .827&6.02 &.809&6.85  &.853&5.80 & .610&21.91&.809&6.85  \\
				&R Kidney 		&-&-  & -&- &-&-  & .966&1.46 & .945&6.80&.837&7.10 &.942&3.88  \\
				&L Kidney 		&-&-  & -&- &-&-  & .959&2.29  &.941&6.83 &.872&7.64 &.948&4.41 \\
				\cline{2-16}
				&mean 		&\textcolor{red}{.965}&5.51  & .959&11.11  &  .915&9.67& .931&3.81  & .933&5.73&.824&10.85 &.928&3.76 \\
				\hline
				\multirow{6}{2cm}{\centering $MargExc-$\\$IL_{swin}$}	    &Liver 			&.965&3.60  &.962&7.33   &.959&8.08  &.958&9.04 &.970&3.46 &.950&4.08 & .974&2.42  \\
				&Spleen 		&-&-  &.961&1.24   &.964&1.18  &.963&1.15 &.953&3.50 &.882&6.00 & .966&1.58  \\
				&Pancreas 		&-&-  &-&-   &.826&5.21  &.816&5.65 &.847&6.24 &.640&31.31 &.817&5.71   \\
				&R Kidney 		&-&-  &-&-   &-&-  &.964&3.49 &.942&7.31 &.883&6.43 &.941&4.63   \\
				&L Kidney 		&-&-  &-&-   &-&-  &.953&2.59 &.933&8.94 &.867&8.96 &.951&3.60   \\
				\cline{2-16}
				&mean 		& \textcolor{red}{.965}&\textcolor{red}{3.60} &\textcolor{red}{.962}&\textcolor{red}{4.28}   &.916&4.821  &.931&4.38   &.929&5.89 &.844&11.36 &.930&3.59 \\
				\hline\hline
				\multirow{6}{2cm}{\centering $MargExcIL$\\$(woMem)$}&Liver 			& .967&5.89 & .965&14.99  & .962&17.45 & .965&6.32 & .972&4.72 & .948& 4.66 & .979&1.83 \\
				&Spleen 	      	& -&- & .963&1.21  & .956&2.42 &.963&1.17   &.957&12.39 &.885&23.77 &.971&1.19 \\
				&Pancreas 	       	& -&- & -&-  & .840&5.72 & .836&5.67 &.865&5.21 &.687&17.62 &.838&5.48  \\
				&R Kidney 	       	& -&- & -&-  & -&- & .968&1.63 & .945&5.44 &.870&6.94 & .937&3.71 \\
				&L Kidney 	        & -&- & -&-  & -&- & .957&2.51 & .941&5.98 & .875&4.22 & .944&3.74  \\
				\cline{2-16}
				&mean 		& \textbf{.967}&5.89 & \textbf{.964}&8.10   & .919&8.53 & \textcolor{red}{.938}&\textbf{3.46}  & \textbf{.936}&6.75&\textcolor{red}{.853}&11.44 & \textcolor{red}{.934}&\textcolor{red}{3.19}\\				
				\hline
				\multirow{6}{2cm}{\centering $MargExcIL$\\$(Ours)$}&Liver 		& .967&2.83 & .966&3.41  & .966&7.05 & .965&7.98 & .971&3.27 & .947&4.85 & .978&1.86 \\
				&Spleen 	    & -&- &  .962&1.18 & .962&1.21 &.962&1.30  & .956&4.60 &.888&6.27 &.969&1.26  \\
				&Pancreas 	       	&-&-  &-&-   & .837&5.32 & .835&5.51 &.865&5.13 &.711&16.78 & .839&5.14 \\
				&R Kidney 	       	&-&-  & -&-  &-&-  & .968&1.40  &.946&6.14 & .846&7.44&.943&3.48 \\
				&L Kidney 	        &-&-  &-&-   &-&-  & .959&2.37  &.942&5.86 & .872&12.20&.935&3.05  \\
				\cline{2-16}
				&mean 		& \textbf{.967}&\textbf{2.83} & \textbf{.964}&\textbf{2.30}   & \textcolor{red}{.922}&\textcolor{red}{4.53} & \textcolor{red}{.938}&3.71  & \textbf{.936}&\textcolor{red}{5.00}&\textcolor{red}{.853}&\textbf{9.51} & \textbf{.935}&\textbf{3.05}\\
				\hline\hline
				\multirow{6}{2cm}{\centering $MargExc$~\cite{ShiMargExc}\\$(Upper$\\$ Bound)$}&Liver & .967&5.89 & .966&6.90  & .968&2.79 & .962&7.01 &.965&3.04 &.952&4.05 & .981&1.63 \\
				&Spleen 	    &-&-  &.950&5.86   &.959&2.15  &  .965&1.15  &.948&3.01 & .896&9.24& .970&1.26\\
				&Pancreas 	       	&-&-  & -&-  & .841&5.92 &  .848&4.83 &.862&5.67 & .677&20.41& .849&4.94 \\
				&R Kidney 	       	&-&-  & -&-  & -&- & .969&1.39&  .950&2.17 &.854&6.65 & .918&7.84 \\
				&L Kidney 	        &-&-  & -&- & -&- & .965&3.96 &  .943&2.28 &.898&10.64 &.935&6.39 \\
				\cline{2-16}
				&mean 				&\textbf{.967}&5.89  & .958&6.38  &  \textbf{.923}&\textbf{3.62}& \textbf{.942}&\textcolor{red}{3.67} &\textcolor{red}{.934}&\textbf{3.23}&\textbf{.855}&\textcolor{red}{10.20} & .931&4.41\\
				\hline
			\end{tabular}
		\end{lrbox}
		\scalebox{0.62}[0.62]{\usebox{\tablebox}}
	\end{table}
	
	\section{Conclusion}
	To unleash the potential from a collection of partially labeled datasets and to settle the efficiency, storage, and ethical issues in current methods, we introduce an incremental learning (IL) mechanism with a practical \textit{four}-stage setting and verify the implementation potential of IL in MOS. 
	IL methods have a natural adaptability to medical image scenarios due to the relatively fixed anatomical structure of human body. The introduced light memory module and loss functions can also stabilize the IL system in practice via constraining the representation of different categories in feature space. 
	We believe that IL holds a great promise in addressing the challenges in real clinics. 

	\bibliographystyle{splncs04}
	\bibliography{Manuscript_IL}

\end{document}